\documentclass[aps,prd,notitlepage,nofootinbib,showpacs,superscriptaddress]{revtex4-1}
\usepackage{graphicx}
\usepackage{amsfonts}
\usepackage{amsmath,amsthm, amsfonts,amssymb}
\usepackage{mathrsfs}
\usepackage{hyperref}
\usepackage{bm}
\usepackage{color}
\usepackage{epstopdf}
\usepackage{epsfig}
\usepackage{algorithm}
\usepackage{algorithmic}

{\rm }

\def\be{\begin{equation}}
 \def\ee{\end{equation}}
 \def\bea{\begin{eqnarray}}
 \def\eea{\end{eqnarray}}
 \def\bes{\begin{eqnarray}}
 \def\ees{\end{eqnarray}}
 \def\bi{\begin{itemize}}
 \def\ei{\end{itemize}} 

\renewcommand{\sec}[1]{\hyperref[sec:#1]{Sec.~\ref{sec:#1}}}

\newcommand{\fig}[1]{\hyperref[fig:#1]{Fig.~\ref{fig:#1}}}

\def\2{\frac{1}{2}}
\def\4{\frac{1}{4}}

\begin{document}

\title{Covert Quantum Internet}

\author{Kamil Br\'adler}
\email{kbradler@uottawa.ca}
\affiliation{Department of Mathematics and Statistics, University of Ottawa, Ottawa, Canada}

\author{George Siopsis}
\email{siopsis@tennessee.edu}
\affiliation{Department of Physics and Astronomy, The University of Tennessee, Knoxville, TN 37996-1200, U.S.A.}

\author{Alex Wozniakowski}
\email{airwozz@gmail.com}
\affiliation{Department of Physics, Harvard University, Cambridge, MA 02138, U.S.A.}

\date{\today}

\begin{abstract}
We apply covert quantum communication based on entanglement generated from the Minkowski vacuum to the setting of quantum computation and quantum networks. Our approach hides the generation and distribution of entanglement in quantum networks by taking advantage of relativistic quantum effects. We devise a suite of covert quantum teleportation protocols that utilize the shared entanglement, local operations, and covert classical communication to transfer or process quantum information in stealth. As an application of our covert suite, we construct two prominent examples of measurement-based quantum computation, namely the teleportation-based quantum computer and the one-way quantum computer. In the latter case we explore the covert generation of graph states, and subsequently outline a protocol for the covert implementation of universal blind quantum computation.
\end{abstract}

\maketitle

\onecolumngrid
\section{Introduction}

The Internet is ubiquitous in daily life, linking a multitude of devices; but the security of the Internet is a public concern. For the exchange of sensitive information a different type of network, called the Darknet, provides anonymous connections that are strongly resistant to eavesdropping and traffic analysis \cite{Syverson et al}. Recently, the structure and resilience of the Internet and the Darknet were analyzed, and the latter was shown to be a more robust network under various types of failures \cite{Domenico and Arenas}. Covert protocols for classical communication \cite{lee2015achieving, sobers2016covert, mukherjee2016covert} and computation \cite{von Ahn et al, Chandran et al, Jarecki} supply networks with a concealing medium or object, enabling data to be transferred or processed without detection. Quantum information science provides a new perspective on the networking of devices, as well as the possible types of algorithms and protocols. In the setting of the quantum internet the network nodes generate, process, and store quantum information locally, and entanglement is distributed across the entire network \cite{Cirac et al 1, H. Jeff Kimble, Rod Van Meter}. As part of the paradigm of local operations and classical communication (LOCC), quantum teleportation protocols utilize the shared entanglement to faithfully transfer quantum data from site to site or implement quantum logic gates for distributed quantum computation \cite{Cirac et al 2, Pirandola et al, Pirandola and Braunstein, Van Meter and Devitt}.

In certain situations, such as a secret government mission, quantum networks might need their communications or computations to be performed without detection. Here, we solve this problem by extending recent work on covert quantum communication \cite{Absolutely Covert} to the setting of quantum computation and quantum networks. We devise a suite of covert quantum teleportation protocols taking advantage of relativistic quantum effects. We show, for instance, that quantum teleportation protocols can be performed in stealth by utilizing the entanglement present in Minkowski vacuum and covert classical communication to hide the non-local parts in quantum teleportation.

In our setup we hide the generation and sharing of entanglement, without a concealing medium or object, between network nodes through the relativistic modes of the Minkowski vacuum, and covert entanglement persists at long distance. We introduce a Minkowski vacuum-assisted amplification scheme in Section \ref{Sect:CovertCommunication}, followed by standard entanglement distillation with covert classical communication to recover covert Bell states. The entanglement-swapping protocol with covert classical communication distributes entanglement across the network, which provides a resource for our suite of covert quantum teleportation protocols. This enables quantum data to be transferred or processed  in the quantum network without detection. Thereby, the quantum network's operations remain hidden from any adversary outside of the network. We call this the \textit{Covert Quantum Internet}.

We apply our suite of covert quantum teleportation protocols to construct two prominent examples of measurement-based quantum computation \cite{Jozsa, Briegel et al}:\ the teleportation-based quantum computer \cite{Gottesman and Chuang, Zhou et al, Eisert et al, Nielsen, Leung, Childs et al} and the one-way quantum computer $\mbox{QC}_{\mathcal{C}}$ \cite{Raussendorf and Briegel 1, Raussendorf and Briegel 2, Raussendorf et al}. Teleportation-based quantum computers use the idea of gate teleportation to carry out computations. We give covert versions of gate teleportation protocols, such as the one-bit teleportation primitive \cite{Zhou et al} and the multipartite compressed teleportation (MCT) protocol \cite{MCT}, in Section \ref{Sect:CovertTeleportation}. We discuss universal quantum computation with covert gate teleportation in Section \ref{Sect:CovertTeleportationComputer}. A one-way quantum computer $\mbox{QC}_{\mathcal{C}}$ carries out computations solely by performing single-qubit measurements on a fixed, many-body resource state and the measurement bases determine the gate or algorithm that is implemented. We show how to covertly generate graph states, such as the topological $3$D cluster state \cite{Raussendorf and Harrington, Raussendorf Harrington Goyal} or the recently discovered Union Jack state with nontrivial $2$D symmetry protected topological order (SPTO) \cite{Miller and Miyake}, in Section \ref{Sect:OneWay}. The single-qubit measurements on the covert resource state are performed locally during implementation of a quantum algorithm, and the feed-forward of measurement outcomes is shared through covert classical communication. In addition we outline a covert implementation of the Broadbent, Fitzsimons, and Kashefi (BFK) protocol for universal blind quantum computation \cite{BFK} in Section \ref{Sect:OneWay}. Finally, we conclude in Section \ref{Sect:Conclusion}.

\section{Covert communication}
\label{Sect:CovertCommunication}

Long before the development of cryptography and encryption, one of humankind’s best techniques of secretly delivering a message was to hide the fact that a message was even delivered at all. Methods for covert quantum communication with a concealing medium, i.e., thermal noise, were considered in \cite{Bash et al, Arrazola and Scarani}, but the conclusion was that stealth capabilities vanish when the medium is absent. In \cite{Absolutely Covert} truly ultimate limits on covert quantum communication were presented. In this section we describe the covert generation and sharing of entanglement between Alice and Bob, or network nodes in a quantum network, without a concealing medium or object using plain two-state inertial detectors \cite{bibRez}. The goal is to generate entanglement from the vacuum. We describe a process which results in entangled detectors. However, the amount of entanglement is very small. We amplify it by repeating the process, thus accumulating a finite amount of entanglement from the vacuum. The fidelity of the resulting state shared by the detectors possessed by Alice and Bob is then increased close to perfection by standard distillation techniques. The previous work of some of the authors,~\cite{Absolutely Covert}, differs from the current analysis in the type of the detectors used. In both cases the detectors are inertial, but in~\cite{Absolutely Covert} we used a pair of two two-level atoms with a time-dependent energy gap. This, however, poses a considerable challenge for the current quantum technology state-of-the-art. Here, the detectors are standard two-level atoms.

Alice and Bob are in possession of two-state detectors.
The detectors couple to vacuum modes which we model by a real massless scalar field $\phi$.
They are separated by a distance $L$. For definiteness, let $\mathbf{r} = \mathbf{r}_A \equiv (0,0,0)$ for Alice, and $\mathbf{r} = \mathbf{r}_B \equiv (0,0,L)$ for Bob. If Alice's (Bob's) detector is turned on at time $t$ ($t'$), then the proper time between the two detectors while in operation is
\be \Delta s^2 = (t'-t)^2 - L^2. \ee
We expect maximal correlations between the two detectors along the null line connecting the two events, i.e., $\Delta s = 0$, or $t'-t= L$. Thus, if Alice's detector is turned on at $t = 0$, then optimally, Bob's detector will be turned on at time $t' = L$.

The two-point correlator for the massless scalar field is given by
\be \Delta (t, \mathbf{r}; t' , \mathbf{r}') = - \frac{1}{4\pi^2}\, \frac{1}{(t-t' - i\varepsilon)^2 - (\mathbf{r}-\mathbf{r}')^2} \; \; .\ee
For best results at finite $L$, for the detectors we choose window functions centered around points at which $\Delta$ diverges (i.e., Alice and Bob are along a null line). Thus, for Alice and Bob, we choose, respectively,
\be\label{eq3} w_A(t) = \lambda \mathbf{w}(t) \ , \ \ w_B(t) = \lambda \mathbf{w}(t-L) \ , \ \ \mathbf{w}(t) = e^{-\frac{t^2}{\sigma^2}}~, \ee
where $\lambda >0$ is a coupling constant whose value depends on the details of the detector setup. In general, it is expected to be small. $\sigma$ is the width of the time window during which the detector is turned on.

Assuming localized detectors, the Hamiltonian is
\be H = H_A + H_B \ , \ \ H_{k} = w_{k} (t) \left( e^{i\delta t} \sigma_k^+ + e^{-i\delta t} \sigma_k^- \right) \phi (t, \mathbf{r}_{k}) \ \ \ (k=A,B) \ee
where $\hbar\delta$ is the energy gap of the two states of a detector, and $\sigma_A^\pm$ ($\sigma_B^\pm$) are spin ladder operators for Alice's (Bob's) detector.

The massless scalar field can be expanded in terms of creation and annihilation operators as
\be \phi (t,\mathbf{r}) = \int \frac{d^3k}{(2\pi)^3 2|\mathbf{k}|} \left( a(\mathbf{k}) e^{-i(|\mathbf{k}| t - \mathbf{k}\cdot \mathbf{r})} + a^\dagger (\mathbf{k}) e^{i(|\mathbf{k}| t - \mathbf{k}\cdot \mathbf{r})}\right)
\ee
where $a(\mathbf{k})$ annihilates the vacuum ($a(\mathbf{k}) |0\rangle = 0$). The Hilbert space of the system is the tensor product of excitations of the vacuum state $|0\rangle$, and the two-dimensional spaces of Alice and Bob spanned by $\{ |0\rangle_k , |1\rangle_k \}$, $k=A,B$.

Assuming an initial state $|in\rangle = |0\rangle\otimes |0\rangle_A \otimes |0\rangle_B$, the evolution of the system is governed by the Hamiltonian $H$. After tracing out the field degrees of freedom, it is easy to see that the final state is of the form
\be\label{eq5} \varrho_{AB} = \left[ \begin{array}{cccc}
a_1 & 0 & 0 & c_1 \\ 0 & a_2 & c_2 & 0 \\ 0 & c_2^\ast & b_2 & 0 \\ c_1^\ast & 0 & 0 & b_1
\end{array}\right] \ee
Thus the evolution of the system of the two detectors after they are switched on and off with the switching profile \eqref{eq3} is given by the quantum channel $\mathcal{N} : |in\rangle \mapsto \varrho_{AB}$.
The final state is in general entangled. However the amount of entanglement is minute. To enhance the entanglement, after the detectors have been switched off and therefore decoupled from the scalar field, we bring the state of $\phi$ back to the vacuum, and repeat the process. This can be repeated $N$ times, where $N$ is large enough for appreciable entanglement generation. At the $n$th step, we apply the channel
\be \mathcal{N}_n : \varrho_{AB}^{(n)} \mapsto \varrho_{AB}^{(n+1)} \ee
At each step, the state is of the same form as \eqref{eq5},
\be \varrho_{AB}^{(n)} = \left[ \begin{array}{cccc}
	a_1^{(n)} & 0 & 0 & c_1^{(n)} \\ 0 & a_2^{(n)} & c_2^{(n)} & 0 \\ 0 & c_2^{(n)\ast} & b_2^{(n)} & 0 \\ c_1^{(n)\ast} & 0 & 0 & b_1^{(n)}
\end{array}\right] \ee
with $\varrho_{AB}^{(0)} = |00\rangle_{AB}\langle 00|$, and $\varrho_{AB}^{(1)} = \varrho_{AB}$.

Using perturbation theory, we obtain
\be \label{eq:iter1storder}
\varrho_{AB} = \left[ \begin{array}{cccc}
	1 & 0 & 0 & 0 \\ 0 & 0 & 0 & 0 \\ 0 & 0 & 0 & 0 \\ 0 & 0 & 0 & 0
\end{array}\right] + \lambda^2 \left[ \begin{array}{cccc}
-2 \mathcal{J}_1 & 0 & 0 & - \mathcal{J}^*_2 \\ 0 & \mathcal{J}_1 & \mathcal{J}_3 & 0 \\ 0 & \mathcal{J}_3^\ast & \mathcal{J}_1 & 0 \\ -\mathcal{J}_2 & 0 & 0 & 0
\end{array}\right] + \mathcal{O} (\lambda^4) 
\ee
Explicitly, at first order in $\lambda^2$ we have
\bea \mathcal{J}_1 &=& - \frac{1}{4\pi^2} \int_{-\infty}^{\infty} dt \mathbf{w}(t)
\int_{-\infty}^{\infty} dt'\mathbf{w}(t')  \frac{e^{i \delta (t -t')}}{(t'-t-i\epsilon)^2  } \, , \nonumber\\
\mathcal{J}_{2} &=& - \frac{1}{4\pi^2} \int_{-\infty}^{\infty} dt \mathbf{w}(t)
\int_{-\infty}^{\infty} dt'\mathbf{w} (t' -L)
\frac{e^{i \delta (t +t')}}{(t'-t-i\epsilon)^2  -L^2} \, ,\nonumber\\
\mathcal{J}_{3} &=& - \frac{1}{4\pi^2} \int_{-\infty}^{\infty} dt \mathbf{w}(t)
\int_{-\infty}^{\infty} dt'\mathbf{w} (t' -L)
\frac{e^{i \delta (t -t')}}{(t'-t-i\epsilon)^2  -L^2} .\eea
These expressions contain poles. Using the Fourier transform of the detector profile,
\be \widetilde{\mathbf{w}} (\omega) = \int_{-\infty}^\infty dt e^{-i\omega t}\mathbf{w}(t) = \sigma\sqrt{\pi} e^{-\sigma^2 \omega^2/4}\ee
and contour integration, we arrive at expressions which are free of singularities and amenable to numerical integration\footnote{The expressions can be evaluated analytically as well.}:
\bea \mathcal{J}_1 &=& \frac{1}{4\pi^2} \int_0^\infty   d\omega\omega  \widetilde{\mathbf{w}}^2 (\omega +\delta), \nonumber\\
\mathcal{J}_{2} &=& \frac{1}{4\pi^2L} \int_0^\infty   d\omega \widetilde{\mathbf{w}}(\omega -\delta)\widetilde{\mathbf{w}}(\omega+\delta)  e^{-iL(\omega-\delta)} \sin \omega L,\nonumber\\
\mathcal{J}_{3} &=& \frac{1}{4\pi^2L} \int_0^\infty   d\omega \widetilde{\mathbf{w}}^2(\omega+\delta)  e^{-iL(\omega-\delta)} \sin \omega L. \eea
After $n$ applications of the quantum channel, we bring the system of the two detectors to the state
\be\label{eq:firstorder} \varrho_{AB}^{(n)} = \left[ \begin{array}{cccc}
	1 & 0 & 0 & 0 \\ 0 & 0 & 0 & 0 \\ 0 & 0 & 0 & 0 \\ 0 & 0 & 0 & 0
\end{array}\right] + n \lambda^2 \left[ \begin{array}{cccc}
	-2 \mathcal{J}_1 & 0 & 0 & - \mathcal{J}^*_2 \\ 0 & \mathcal{J}_1 & \mathcal{J}_3 & 0 \\ 0 & \mathcal{J}_3^\ast & \mathcal{J}_1 & 0 \\ -\mathcal{J}_2 & 0 & 0 & 0
\end{array}\right] + \mathcal{O} (\lambda^4) \ee
Thus, after $N$ steps, the effective coupling constant becomes $N\lambda^2$, and entanglement is amplified. To quantify the results, we
use the singlet fraction of $\varrho$ defined as~\cite{horodeckisReview}
\be\label{eq7} \mathcal{F} (\varrho) = \max_\Phi \langle \Phi |\varrho |\Phi \rangle \ , \ \ |\Phi \rangle =  U_A\otimes U_B \frac{1}{\sqrt{2}}\left( |00\rangle_{AB} + |11\rangle_{AB} \right).
\ee
\begin{figure}[h]
	\centering
	\includegraphics[width=14.5cm]{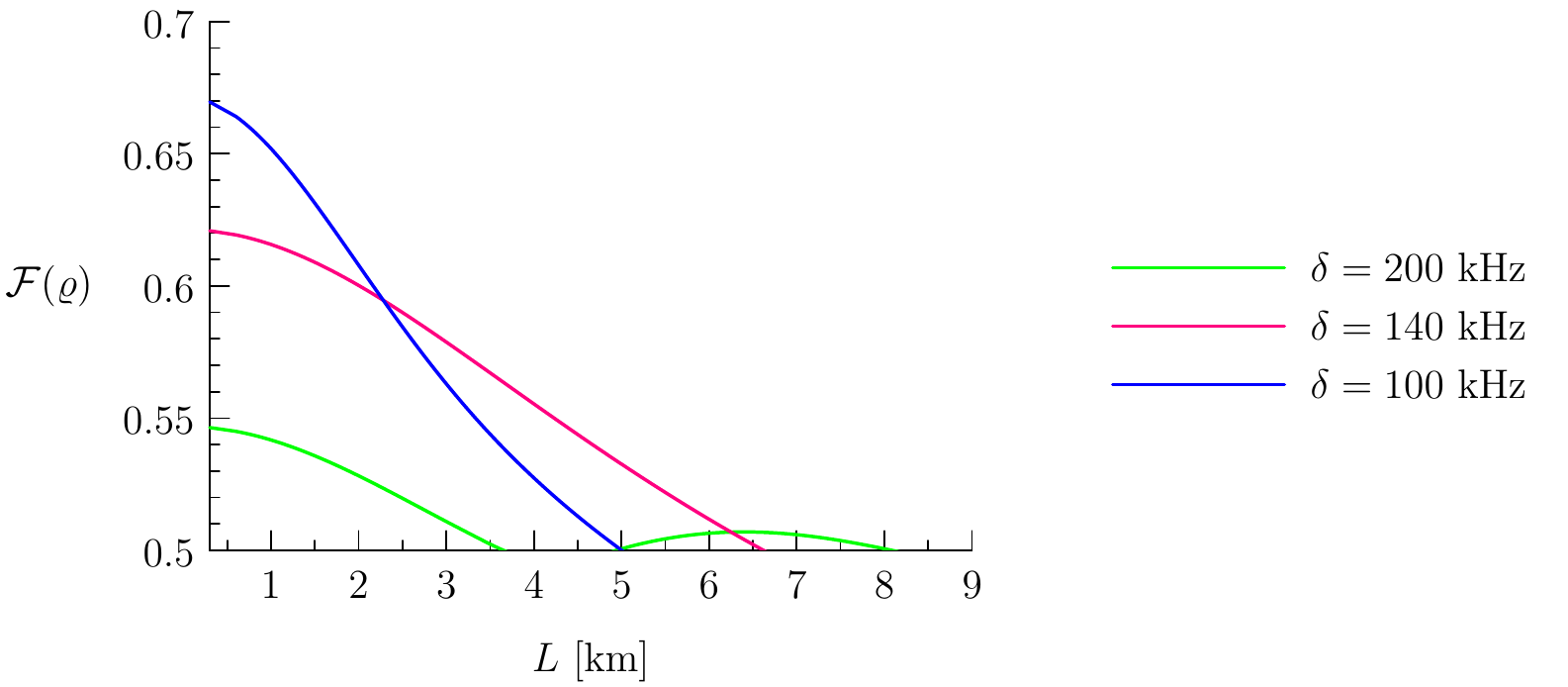}
	\caption{\label{fig:1} The singlet fraction \eqref{eq7} \emph{vs.}\ the distance $L$ of the two detectors for different values of their energy gap. The number of iterations is $N=500,\lambda^2=0.01$, and $\sigma=10~\mu\mathrm{s}$.}
\end{figure}
 \begin{figure}[h]
	\centering
	\includegraphics[width=14.5cm]{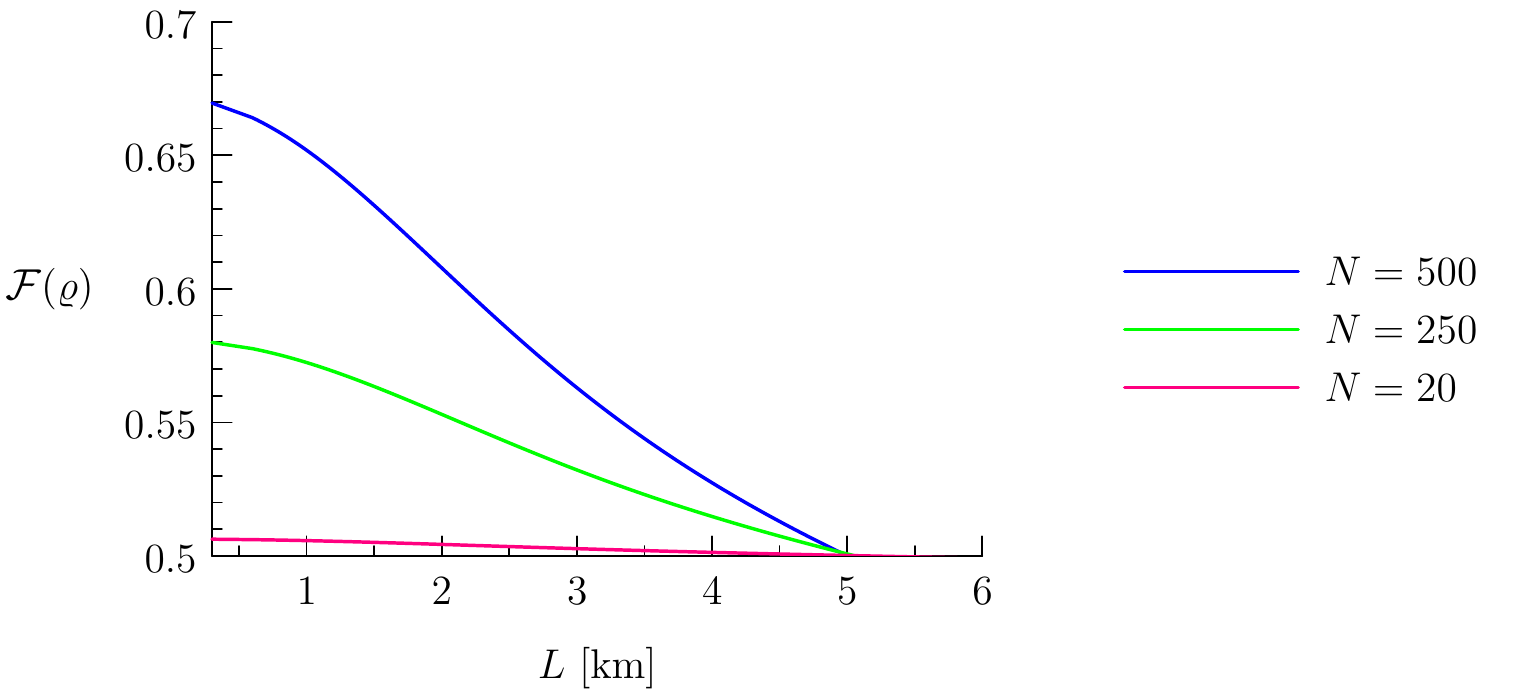}
	\caption{\label{fig:2} The singlet fraction \eqref{eq7} \emph{vs.}\ the distance $L$ of the two detectors  at different stages of entanglement amplification for $\lambda^2=0.01,\delta=100~\mathrm{kHz}$, and $\sigma=10~\mu\mathrm{s}$.}
\end{figure}
We have optimized the singlet fraction using perturbation theory numerically to second order in $\lambda^2$. We have shown above explicit analytic expressions at first order in $\lambda^2$. We have also obtained second-order analytical expressions, but not included here for brevity. Their derivation is based on the expansion of two two-level detectors to an arbitrary perturbative order in~\cite{bradlerExpansion}. Using these second-order analytical expressions, we have calculated the singlet fraction \eqref{eq7} numerically. Its behavior is shown in Fig.~\ref{fig:1} and ~\ref{fig:2}. For a wide range of parameters of the system, we obtain $\mathcal{F} > 1/2$, showing entanglement extraction from the vacuum. More precisely, there are known two-qubit entangled states where $\mathcal{F}<1/2$, but in our case the entanglement of formation (EOF)~\cite{horodeckisReview} is zero whenever $\mathcal{F}\leq1/2$, and positive otherwise. The iteratively collected entanglement through the Minkowski vacuum-assisted amplification scheme is shown in figure Fig.~\ref{fig:2}. As the number of iterations increases, the original minute amount of entanglement reaches quite substantial values. When measured as the singlet fraction, it goes well above $1/2$ and a corresponding increase is documented for the EOF as well. Recall that the EOF is a true entanglement measure for two qubits. As in the case of Fig.~\ref{fig:1}, the actual calculation is done to the second order in $\lambda^2$. After the amplification process is completed, we perform standard distillation on many copies of the iterated state in order to arrive at a nearly perfect maximally entangled state. Thus, we produce a covert Bell state shared between Alice's and Bob's detectors.

\section{Covert Quantum Teleportation}
\label{Sect:CovertTeleportation}

The quantum teleportation protocol was introduced by Bennett et al.\ \cite{Bennett et al}. Recently, the protocol was found to have a $3$D topological structure using the quon language \cite{Quon}; and Pirandola and Braunstein cite teleportation as the ``most promising mechanism for a future quantum internet" \cite{Pirandola and Braunstein}. To implement the protocol for two parties: a sender, Alice, disassembles an unknown quantum state at her location; and a receiver, Bob, reconstructs the quantum state identically at his location. In order for the reconstruction to succeed, Alice and Bob prearrange to share the Bell state, which is utilized as a resource for the protocol. In addition, the parties share some purely classical information.

The idea of covertly implementing the quantum teleportation protocol involves hiding the non-local parts, i.e., distribution of the entangled resource state and classical communication of measurement outcomes. In \ref{Sect:CovertCommunication} we presented the covert generation and distribution of the Bell state, shared by Alice and Bob. Covert classical communication hides the transfer of the measurement outcomes from Alice to Bob. Thus, we establish a protocol for the hidden transfer of quantum data between Alice and Bob, or network nodes.

Gottesman and Chuang considered a variant of the teleportation protocol in which Bob's reconstructed quantum state differed from Alice's original quantum state \cite{Gottesman and Chuang}. In this elegant version of teleportation the fault-tolerant construction of certain quantum gates was developed. Later, Zhou et al.\ extended the teleportation method of gate construction by introducing the one-bit teleportation primitive, which enabled a class of gates in the Clifford hierarchy to be recursively constructed \cite{Zhou et al}. As an example, this includes the controlled rotations that appear in Shor's factoring algorithm \cite{Shor}. Zhou et al.\ and Eisert et al.\  first considered the minimal resources required for the implementation of certain remote quantum gates \cite{Zhou et al, Eisert et al}. A generalization of these methods, called the multipartite compressed teleportation (MCT) protocol, enables the efficient sharing of multipartite, non-local quantum gates in which the protocol does not reduce to compositions of bipartite teleportation \cite{MCT}. Furthermore, the MCT protocol allows a quantum network to share a controlled gate with multiple targets.

The scheme for the covert teleportation of quantum states is adaptable to the variants of the protocol that teleport quantum gates.  For instance, the covert teleportation of a controlled-NOT gate utilizes one covert Bell state, local operations, and covert classical communication of measurement outcomes. The covert recursive construction with one bit teleportation \cite{Zhou et al} utilizes instances of the covert, controlled-NOT gate, local operations, and covert classical communication of the ancillary state preparation and measurement outcomes. In the MCT protocol the resource state is either the Greenberger-Horne-Zeilinger state $|\mbox{GHZ} \rangle$, or $|\mbox{Max} \rangle$ as first introduced in \cite{Max}, depending upon the particular non-local quantum gate \cite{MCT}. Both multipartite resource states can be constructed from simpler Bell states \cite{Bose et al, Max}, thus covert Bell states can be distilled into covert $|\mbox{GHZ} \rangle$ or $|\mbox{Max} \rangle$ with local operations and covert classical communication.

\section{Covert Measurement-Based Quantum Computation}

\subsection{Teleportation-Based Quantum Computation}
\label{Sect:CovertTeleportationComputer}

The teleportation-based approach to quantum computation uses the idea of gate teleportation to affect quantum computation \cite{Gottesman and Chuang, Zhou et al, Eisert et al, Nielsen, Leung, Childs et al, MCT}. Given the ability to perform single-qubit gates, the teleportation of either a controlled-NOT gate or a controlled-Z gate is universal for quantum computation. Hence, a covert implementation of a universal teleportation-based quantum computer is achieved with covert Bell states, local operations that include all single-qubit gates, and covert classical communication of measurement outcomes. Alternative universal quantum gate sets can be constructed through the covert MCT protocol or the covert recursive construction with the one-bit teleportation primitive.

\subsection{One-Way Quantum Computation}
\label{Sect:OneWay}

Another prominent example of measurement-based quantum computation is the one-way quantum computer $\mbox{QC}_{\mathcal{C}}$, which was introduced by Raussendorf and Briegel \cite{Raussendorf and Briegel 1, Raussendorf and Briegel 2}. The idea of $\mbox{QC}_{\mathcal{C}}$ is to prepare a fixed, many-body resource state in which quantum computations are carried out solely through single-qubit measurements on the resource state and classical feed-forward of measurement outcomes \cite{Jozsa, Briegel et al}. Cluster states, a sub-class of graph states, are the archetypal resource state, whereby topological $3$D cluster states are utilized in fault-tolerant versions of $\mbox{QC}_{\mathcal{C}}$ \cite{Raussendorf and Harrington, Raussendorf Harrington Goyal}. Universal resource states for $\mbox{QC}_{\mathcal{C}}$ have been widely studied \cite{Briegel et al, Raussendorf and Harrington, Raussendorf Harrington Goyal, BFK, Van den Nest et al}, and recently the universal Union Jack state with nontrivial $2$D symmetry protected topological order (SPTO) was found \cite{Miller and Miyake}.

A one-way quantum computation proceeds with a classical input that specifies the data and program. A graph state is generated by preparing each vertex qubit in a graph in the fiducial starting state $| + \rangle \equiv \frac{1}{\sqrt{2}} (|0 \rangle + | 1 \rangle)$, and applying a controlled-Z gate to every pair of qubits connected by an edge in the graph. Since controlled-Z gates mutually commute, the production of a graph state is independent of the order of operations. Next, a sequence of adaptive single-qubit measurements is implemented on certain qubits in the graph, whereby the  measurement bases depend upon the specified program as well as the previous measurement outcomes. A classical computer determines which measurement directions are chosen during every step of the computation.

To covertly implement $\mbox{QC}_{\mathcal{C}}$ the classical input that specifies the data and program needs to be shared by covert classical communication. The graph state is generated by applying the covert, controlled-Z gate to each pair of fiducial qubits connected by an edge in the graph. The single-qubit measurements are applied locally. The classical computer that determines the measurement directions is offline, and covert protocols for communication \cite{lee2015achieving, sobers2016covert, mukherjee2016covert} and computation \cite{von Ahn et al, Chandran et al, Jarecki} are utilized.

Topological $3$D cluster states provide a pathway towards large-scale distributed quantum computation \cite{Van Meter and Devitt} by combining the universality of $2$D cluster states with the topological error-correcting capabilities of the toric code \cite{Kitaev}. Quantum computation is performed on a $3$D cluster state via a temporal sequence of single-qubit measurements, which leaves a non-trivial cluster topology that embeds a fault-tolerant quantum circuit \cite{Raussendorf and Harrington, Raussendorf Harrington Goyal}. To generate the computational resource state fiducial qubits are located at the center of faces and edges of an elementary cell; see Figure 2, page $5$ of \cite{Raussendorf Harrington Goyal}. Controlled-Z gates are applied from each face qubit to each neighboring edge qubit. The elementary cell is tiled in $3$D to form a topological cluster state. Hence, the covert generation of the topological $3$D cluster state follows from the discussion above.

The universal $2$D cluster state was shown to have trivial $2$D symmetry protected topological order (SPTO) and nontrivial $1$D SPTO, whereas the universal Union Jack state was found to possess nontrivial $2$D SPTO \cite{Miller and Miyake}. The Union Jack state is generated by preparing fiducial qubits, and applying a controlled-controlled-Z gate to every triangular cell in the graph. The resulting universal resource state has the advantageous property of being Pauli universal, meaning that single-qubit measurements in the Pauli bases on the Union Jack state can implement arbitrary quantum computations. The feature of Pauli universality is forbidden for the $2$D cluster state, as implied by the Gottesman-Knill theorem, since the state is generated by an element from the Clifford group \cite{Gottesman Knill}. In other words single-qubit measurements in the Pauli bases on the $2$D cluster state are efficiently simulated on a classical computer. The covert generation of the Union Jack state is a simple modification of the aforementioned scheme with controlled-Z gates for graph states. Namely, covert teleportation applies a doubly controlled-Z gate to fiducial qubits in triangular cells.

BFK introduced a universal blind quantum computation protocol \cite{BFK}, which exploits measurement-based quantum computation to allow a client, Alice, with limited quantum computing power to delegate a computation to a quantum server, Bob. The premise of the BFK  protocol is that Alice's inputs, computations, and outputs are unknown to Bob \cite{BFK, Chien et al}. In what follows we outline a covert implementation of the basic steps in the BFK protocol, by hiding the communication rounds between Alice and Bob.

\begin{algorithm}[H]
\floatname{algorithm}{Protocol}
\renewcommand{\thealgorithm}{}
\caption{Covert BFK universal blind quantum computation}
\begin{enumerate}
\item The preparation stage \newline
For each column $x=1,\dots,n$ \newline
For each row $y=1,\dots,m$ \newline
\begin{enumerate}
\item Alice prepares qubits $| \psi_{x,y} \rangle$ such that
$$|\psi_{x,y} \rangle \in \Big\{ \frac{1}{\sqrt{2}}  \big (|0 \rangle + e^{i  \theta_{x,y}} |1 \rangle \big) \; | \; \theta_{x,y} = 0, \pi / 4, \dots, 7 \pi / 4 \Big\},$$
and transfers each qubit to Bob via covert teleportation.
\item Bob generates an entangled brickwork state $G_{n \times m}$ \cite{BFK} from all the qubits received, applying a controlled-Z gate to qubits joined by an edge.
\end{enumerate}
\item Interaction and measurement stage \newline
For each column $x=1,\dots,n$ \newline
For each row $y=1,\dots,m$ \newline
\begin{enumerate}
\item Alice computes $\phi'_{x,y}$ based on the real measurement angle $\phi_{x,y}$ and the previous measurement results.
\item Alice chooses $r_{x,y} \in \{0,1\}$ and computes the angle $$\delta_{x,y} = \phi'_{x,y} + \theta_{x,y} + \pi r_{x,y}.$$
\item Alice sends $\delta_{x,y}$ to Bob via covert classical communication.
\item Bob measures $|\psi_{x,y} \rangle$ in the basis $ \{ |0 \rangle + e^{i \delta_{x,y}} | 1 \rangle \} ,$ then sends the measurement outcome to Alice via covert classical communication.
\item If $r_{x,y} = 1$ above, then Alice flips the measurement outcome bit; otherwise she does nothing.
\end{enumerate}
\end{enumerate}
\end{algorithm}

\section{Conclusion}
\label{Sect:Conclusion}

A quantum internet is a manifest platform for quantum cryptography, sensor networks, and large-scale networked quantum computing. We developed a quantum internet that hides its operations by exploiting features of relativistic quantum information. In particular, we used properties of the Minkowski vacuum to hide the generation and distribution of entanglement in quantum networks, which provides a resource for covert quantum teleportation, thus enabling the faithful transmission of quantum data from site to site without detection, as well as the construction of universal quantum computers that carry out quantum logic operations in stealth. We anticipate further protocols being constructed covertly. For instance, graph states can be generated covertly, which could be used in the study of preexisting protocols and algorithms that require graph states. We are working on the experimental details of our setup towards a proof-of-principle demonstration in the near future.

\acknowledgements
This material is based upon work supported by the U.S.\ Air Force Office of Scientific Research under award number FA9550-17-1-0083. The authors also acknowledge support from the U.S.\ Office of Naval Research (ONR) under award number N00014-15-1-2646.

\end{document}